\documentclass[preprint,aps]{revtex4}
\usepackage{euscript}
\usepackage{amsmath}
\usepackage{epsfig}
\begin{document}
\preprint{UCI-TR-2009-02}
\title{Vacuum Tunneling in an Electroweak Model in Extra Dimensions With an External Flux}
\author{Aaron Roy\footnote{Electronic address:roya@uci.edu}
}
\affiliation{
Department of Physics and Astronomy, University of California, Irvine,
California 92697-4575}

\begin{abstract}
With the standard system for an SU(2) Higgs field in M$_4 \, \otimes \, $S$_1$, the top and bottom component of the Higgs spinor have exactly the same coefficients for the quadratic and quadric terms. This makes the vacuum degenerate and thus there are no tunneling effects to zeroth order in radiative corrections of the vector gauge fields in the standard model with this extra dimensional geometry. However, if we include an external magnetic flux that permeates our manifold, then the top component of the Higgs spinor will have an additional term in its coefficient due to this theoretical flux with the usual charge assignments for the standard model. This extra term gives rise to two non-degenerate vacuum states for the resulting potential. We will compute the tunneling probability per unit time per unit volume between these vacuum states for the zero modes of our system as well as investigate the masses for the fields of the model using the false vacuum.     
\end{abstract}

\maketitle

\section{Introduction}
In this paper we will discuss the tunneling between vacuum states in M$_4 \, \otimes \, $S$_1$ [1,2,3] that are no longer degenerate with the inclusion of an external magnetic flux. We will be following the formalism of [4] as well as adopting the same notation, even though our problem is more involved mathematically from that example. The Higgs potential for this paper will get an additional contribution to the upper component of the Higgs spinor that will involve the flux [5]. With the usual charge assignments, only the top component of the spinor is affected by this flux since it is charged, the bottom component is unaffected since it is neutral. This will create a local minimum in our theory as well as the usual global minimum which will be denoted as the true vacuum.

We will compute the tunneling probability per unit time per unit volume from the false vacuum to the true vacuum. This tunneling probability will be a function of the flux for our model. In the limit as the flux goes to zero, the tunneling probability will go to zero as expected, for then the vacuum becomes degenerate. Obviously there is still tunneling between the degenerate vacuums but this is tunneling in a different context than we are analyzing in this paper. A WKB approximative method [4] will be used to determine the predominate portion of the tunneling probability, which involves a decaying exponential. There is also a multiplicative constant that is a parameter [4] to the tunneling probability formula that we will not be concerning ourselves with since it, to my knowledge, can only be determined numerically and does not affect the overall behavior of the tunneling probability for the numbers that will be used in this paper. In Sec II a summary of the general theory [5] will be presented. In Sec. III the masses for the fields in the model using the false vacuum will be determined.  In Sec. IV we will discuss the Higgs potential for the tunneling and the Lagrangian density for the Higgs field. We will also integrate out the extra coordinate by forming the action for the model thus leaving an effective 4-D Lagrangian density in terms of the field modes. In Sec. V the problem of calculating the probability of tunneling for the zero mode (standard model Higgs field) between the false vacuum and the true vacuum will be analyzed. In Sec. VI the general result for the tunneling probability will be presented as well as a discussion of appropriate limits and then a short calculation using the upper limit for the ratio of flux to compactification size from [5].               

\section{Summary of the theory}
Originally from [5],
\begin{equation}
\EuScript{L} = (D_A \varphi )^\dagger (D^A \varphi) - \frac{1}{2}Tr(F_{AB}F^{AB}) - \frac{1}{4}f_{AB}f^{AB} + \mu ^2 \varphi ^\dagger \varphi - \frac{\lambda}{2}(\varphi ^\dagger \varphi )^2
\end{equation}
where $A=0,1,2,3,5$ and as usual $D_A = \partial _A + igW_A + \frac{i}{2}g'B_A$. The metric used was 
\begin{equation}
g^{AB}=
  \begin{cases} 
    0& \text{if $A\neq B$},\\
   -1& \text{if $A=B=1,2,3,5$},\\ 
    1& \text{if $A=B=0$} 
  \end{cases} 
\end{equation}
where $y$ is the extra coordinate on the circle of radius $R$ for our manifold. Then we added an external flux permeating the manifold. This flux only affects the charged fields in the model with an additional phase $e^{iQby/R}$ where $Q$ is the charge of the field and $b = \frac{e}{\hbar c}\times {\rm flux}$ (Gaussian units). We will be using 
($\hbar = c = 1$). Given the usual charge assignments for the standard model we found $\phi \underset{\rm{flux}}{\longrightarrow} \begin{pmatrix} e^{iby/R} & 0 \\ 0 & 1\end{pmatrix}\phi$ or
\begin{equation}
\phi \underset{\rm{flux}}{\longrightarrow}B\phi 
\end{equation}
where $B = \begin{pmatrix} e^{iby/R} & 0 \\ 0 & 1\end{pmatrix}$. Similarly 
$W_A = \begin{pmatrix} \frac{1}{2}W_A ^3 & \frac{1}{\sqrt{2}}W_A ^+ \\ \frac{1}{\sqrt{2}}W_A ^- & - \frac{1}{2}W_A ^3\end{pmatrix} \underset{\rm{flux}}{\longrightarrow} \\ \begin{pmatrix} \frac{1}{2}W_A ^3 & \frac{1}{\sqrt{2}}e^{iby/R}W_A ^+ \\ \frac{1}{\sqrt{2}}e^{-iby/R}W_A ^- & - \frac{1}{2}W_A ^3\end{pmatrix} = \begin{pmatrix} e^{iby/R} & 0 \\ 0 & 1\end{pmatrix}W_A \begin{pmatrix} e^{-iby/R} & 0 \\ 0 & 1\end{pmatrix}$ or 
\begin{equation}
W_A \underset{\rm{flux}}{\longrightarrow} B W_A B^\dagger .                                   
\end{equation}

For ${\rm{Tr}}(F_{AB}F^{AB}) = {\rm{Tr}}(F_{\mu \nu}F^{\mu \nu}) + 2{\rm{Tr}}(F_{\mu 5}F^{\mu 5})$ it was found that \\
${\rm{Tr}}(F_{\mu \nu}F^{\mu \nu})\underset{\rm{flux}}{\longrightarrow}{\rm{Tr}}(F_{\mu \nu}F^{\mu \nu})$ and 
\begin{equation}
{\rm{Tr}}(F_{\mu 5}F^{\mu 5})\underset{\rm{flux}}{\longrightarrow}\frac{1}{2}F^3 _{\mu 5}F^{3\mu 5} - (\partial _\mu W_5 ^- - \partial _y W_\mu ^- + \frac{ib}{R}W_\mu ^- )(\partial ^\mu W_5 ^+ - \partial _y W^{\mu +} - \frac{ib}{R}W^{\mu +}) \nonumber
\end{equation}
\begin{equation}
+ (\rm{cubic \, and \, quadric \, terms}) 
\end{equation}
where as usual $W_A ^+ = \frac{1}{\sqrt{2}}(W_A ^1 - i W_A ^2)$ and  
$W_A ^- = \frac{1}{\sqrt{2}}(W_A ^1 + i W_A ^2)$. Then with the field definitions,
\begin{eqnarray}
\tilde{W}^- _\mu = W^- _\mu - \Lambda \partial _\mu W_5 ^- \\
\tilde{W}^+ _\mu = W^+ _\mu - \beta \partial _\mu W_5 ^+ \, ,
\end{eqnarray} 
we finally had
\begin{equation}
{\rm{Tr}}(F_{\mu 5}F^{\mu 5})\underset{\rm{flux}}{\longrightarrow}\frac{1}{2}F^3 _{\mu 5}F^{3\mu 5}  - ( \partial _y \tilde{W}_\mu ^- - \frac{ib}{R}\tilde{W}_\mu ^- )(\partial _y \tilde{W}^{\mu +} + \frac{ib}{R}\tilde{W}^{\mu +}) \nonumber
\end{equation}
\begin{equation}
+ (\rm{cubic \, and \, quadric\, terms}) 
\end{equation}
where $\Lambda = (\partial _y - \frac{ib}{R})^{-1}$ and $\beta = (\partial _y + \frac{ib}{R})^{-1}$. The reason that these fields were redefined was to solve the degrees of freedom problem that arises from the charged $W$'s picking up a mass from the flux before any Higgs mechanism [5]. 

Now for the Higgs particle we had 
\begin{equation}
(D_A \phi)^\dagger (D^A \phi) = \phi ^\dagger[\overleftarrow{\partial} _A  + igW_A + \frac{i}{2}g'B_A][\overrightarrow{\partial} ^A - igW^A - \frac{i}{2}g'B^A]\phi \underset{\rm{flux}}{\longrightarrow} \nonumber
\end{equation}
\begin{equation}
\phi ^\dagger B^\dagger[\overleftarrow{\partial} _A + igB W_A B^\dagger + \frac{i}{2}g'B_A][\overrightarrow{\partial} ^A - igB W^A B^\dagger - \frac{i}{2}g'B^A]B\phi \nonumber
\end{equation}
\begin{equation}
= \phi ^\dagger[\overleftarrow{\partial} _\mu + igW_\mu + \frac{i}{2}g'B_\mu][\overrightarrow{\partial} ^\mu - igW^\mu - \frac{i}{2}g'B^\mu]\phi \nonumber
\end{equation}
\begin{equation} 
- \phi ^\dagger[\overleftarrow{\partial _y} + (\partial _y B^\dagger )B + igW_5 + \frac{i}{2}g'B_5][\overrightarrow{\partial _y} + B^\dagger (\partial _y B) - igW_5 - \frac{i}{2}g'B_5]\phi \nonumber
\end{equation}
or in terms of (6) and (7),
\begin{eqnarray} 
 (D_A \phi)^\dagger (D^A \phi)\underset{\rm{flux}}{\longrightarrow}   
\phi ^\dagger[\overleftarrow{\partial} _\mu + ig\tilde{W}_\mu + ig\partial _\mu T + \frac{i}{2}g'B_\mu][\overrightarrow{\partial} ^\mu - ig\tilde{W}^\mu - ig\partial ^\mu T - \frac{i}{2}g'B^\mu]\phi \nonumber\\ 
- \phi ^\dagger[\overleftarrow{\partial _y} + (\partial _y B^\dagger )B + igW_5 + \frac{i}{2}g'B_5][\overrightarrow{\partial _y} + B^\dagger (\partial _y B) - igW_5 - \frac{i}{2}g'B_5]\phi
\end{eqnarray}
where $\tilde{W}_\mu = \begin{pmatrix} \frac{1}{2}W^3 _\mu & \frac{1}{\sqrt{2}}\tilde{W} ^+ _\mu \\ \frac{1}{\sqrt{2}}\tilde{W} ^- _\mu & -\frac{1}{2}W^3 _\mu\end{pmatrix}$ and 
$T = \frac{1}{\sqrt{2}}\begin{pmatrix} 0 & \beta W^+ _5 \\ \Lambda W^- _5 & 0\end{pmatrix}$. Combining the term 
$-\phi ^\dagger (\partial _y B^\dagger)(\partial _y B)\phi = -\phi ^\dagger
\begin{pmatrix} \frac{b^2}{R^2} & 0 \\ 0 & 0\end{pmatrix}\phi$ in equation (9) with the terms $\mu ^2 \phi ^\dagger \phi - \frac{\lambda}{2}(\phi ^\dagger \phi )^2$, we defined the potential, 
\begin{equation}
V(\phi ^\dagger \phi) = -\phi ^\dagger \begin{pmatrix} \mu ^2 - \frac{b^2}{R^2} & 0 \\ 0 & \mu ^2\end{pmatrix}\phi + 
\frac{\lambda}{2}(\phi ^\dagger \phi )^2.
\end{equation} 
Minimizing the potential in equation (10), the true vacuum is $\langle \phi \rangle _0 = \begin{pmatrix} 0 \\ \sqrt{\frac{\mu ^2}{\lambda}} \end{pmatrix}$ for the global minimum and the false vacuum is $\langle \phi \rangle _{\rm{local \, minimum}} = \begin{pmatrix} \sqrt{\frac{\mu ^2 - \frac{b^2}{R^2}}{\lambda}} \\ 0\end{pmatrix}$. For more details of the general theory please see [5].
\section{The false vacuum}  
We now want to look at the masses for the particles in this theory using the false vacuum. Denoting the false vacuum as 
$\frac{1}{\sqrt{2}}\begin{pmatrix} v \\ 0 \end{pmatrix}$ where $v = \sqrt{\frac{\mu ^2 - \frac{b^2}{R^2}}{\lambda}}$, then from (8) and (9) we find for the masses,
\begin{equation}
m_W = \sqrt{\frac{g^2 (\mu ^2 - \frac{b^2}{R^2})}{4\lambda} + \frac{b^2}{R^2}}
\end{equation}
for $\tilde{W}_\mu ^{+-}$ and 
\begin{equation}
m_Z = \sqrt{\frac{(g^2 + g'^2)(\mu ^2 - \frac{b^2}{R^2})}{4\lambda}}
\end{equation}
for $Z_A$ where $Z_A = \frac{1}{\sqrt{g^2 + g'^2}}(g W^3 _A + g' B_A)$. The photon is massless were \\ 
$A_A = \frac{1}{\sqrt{g^2 + g'^2}}(g' W^3 _A - g B_A)$. Note that the definitions for $Z_A$ and $A_A$ differ from the standard model because of the polarization of the false vacuum. Then the mass for the remaining Higgs field (it will be in the top component now with the false vacuum [5]) is 
\begin{equation}
m_h = \sqrt{4\mu ^2 + \frac{b^2}{R^2}} \, .
\end{equation}
and the mass for $W_5 ^{+-}$ is 
\begin{equation}
m_{W_5} = \frac{g}{2}\sqrt{\frac{\mu ^2 - \frac{b^2}{R^2}}{\lambda}} \, .
\end{equation}
Also
\begin{equation}
\frac{m_W}{m_Z} = \sqrt{\frac{g^2 (\mu ^2 - \frac{b^2}{R^2}) + 4\lambda \frac{b^2}{R^2}}{(g^2 + g^{'2})(\mu ^2 - \frac{b^2}{R^2})}} \, .
\end{equation}

With 
\begin{equation}
-\frac{1}{2}F_{\mu 5}^3 F^{3\mu 5} = \frac{1}{2(g^2 + g'^2)}(g\partial _\mu Z_5 + g'\partial _\mu A_5 - g\partial _y Z_\mu
-g'\partial _y A_\mu )(g\partial ^\mu Z_5 + g'\partial ^\mu A_5 - g\partial _y Z^\mu
-g'\partial _y A^\mu ) \nonumber
\end{equation}
\begin{equation}
+ (\rm{cubic \, and \, quadric \, terms}) \nonumber
\end{equation}
and 
\begin{equation}
-\frac{1}{2}f_{\mu 5}f^{\mu 5} = \frac{1}{2(g^2 + g'^2)}(g\partial _\mu A_5 - g'\partial _\mu Z_5 + g'\partial _y Z_\mu - g\partial _y A_\mu)(g\partial ^\mu A^5 - g'\partial ^\mu Z_5 + g'\partial _y Z^\mu - g\partial _y A^\mu) \nonumber
\end{equation} 
along with (9), we have for the modes
\begin{equation}
m_{W_n} = \sqrt{\frac{g^2 (\mu ^2 - \frac{b^2}{R^2})}{4\lambda} + \frac{(n+b)^2}{R^2}}
\end{equation}
and 
\begin{equation}
m_{Z_n} = \sqrt{\frac{(g^2 + g'^2)(\mu ^2 - \frac{b^2}{R^2})}{4\lambda} + \frac{n^2}{R^2}} \, .
\end{equation}
along with
\begin{equation}
m_{A_n} = \frac{|n|}{R}
\end{equation}
for the photon modes. The mass for the modes of the Higgs field are 
\begin{equation}
m_h = \sqrt{4\mu ^2 + \frac{(n+b)^2}{R^2}}
\end{equation}
and $W_5$ has no mode dependence for its mass [5].

Adding first generation leptons and quarks gives \\
$\EuScript{L}_{\rm{fermions}} = i\begin{pmatrix} \bar{\nu}^e _L , & \bar{e}_L \end{pmatrix}D_\mu \gamma ^\mu \begin{pmatrix} \nu ^e _L \\ e_L \end{pmatrix} + i\bar{e}_R D_\mu \gamma ^\mu e_R + i\begin{pmatrix} \bar{u}_L , & \bar{d}_L \end{pmatrix}D_\mu \gamma ^\mu \begin{pmatrix} u_L \\ d_L \end{pmatrix} \\ + i\bar{u}_R D_\mu \gamma ^\mu u_R +  i\bar{d}_R D_\mu \gamma ^\mu d_R - \begin{pmatrix} \bar{\nu}^e _L , & \bar{e}_L \end{pmatrix}D_5 \gamma ^5 \begin{pmatrix} \nu ^e _R \\ e_R \end{pmatrix} - \begin{pmatrix} \bar{u}_L , & \bar{d}_L \end{pmatrix}D_5 \gamma ^5 \begin{pmatrix} u_R \\ d_R \end{pmatrix} \\ - \lambda _e \begin{pmatrix} \bar{\nu}^e _L , & \bar{e}_L \end{pmatrix}i\tau ^2\phi ^* e_R - \lambda _d \begin{pmatrix} \bar{u}_L , & \bar{d}_L \end{pmatrix}i\tau ^2\phi ^* d_R - \lambda _u \begin{pmatrix} \bar{u}_L , & \bar{d}_L \end{pmatrix}\phi \, u_R$ \, where the same notation is used in [5]. Notice that the Higgs coupling terms violate U(1). This is the price we have to pay in order to get mass terms for the fermions in the model using the false vacuum. Of course as in [5], the $\gamma ^5$ terms violate SU(2) but these terms are necessary if there is to be mode dependence of the masses for the fermion modes. The masses are 
\begin{equation}
m_e = \lambda _e \sqrt{\frac{\mu ^2 - \frac{b^2}{R^2}}{2\lambda}}
\end{equation}  
for the electron and
\begin{equation}
m_u = \lambda _u \sqrt{\frac{\mu ^2 - \frac{b^2}{R^2}}{2\lambda}}
\end{equation}
for the up quark and finally,
\begin{equation}
m_d = \lambda _d \sqrt{\frac{\mu ^2 - \frac{b^2}{R^2}}{2\lambda}}
\end{equation}
for the down quark. 

For the masses of the modes, we have from \\  $-\begin{pmatrix} \bar{\nu}^e _L , & \bar{e}_L \end{pmatrix}[\partial _y + \begin{pmatrix} 0 & 0 \\ 0 & \frac{-ib}{R}\end{pmatrix}]\gamma ^5 \begin{pmatrix} \nu ^e _R \\ e_R \end{pmatrix}$ and $-\begin{pmatrix} \bar{u} _L ,  & \bar{d}_L \end{pmatrix}[\partial _y + \begin{pmatrix} \frac{2}{3}\frac{ib}{R} & 0 \\ 0 & \frac{1}{3}\frac{ib}{R}\end{pmatrix}]\gamma ^5 \begin{pmatrix} u _R \\ d_R \end{pmatrix}$, which gives 
\begin{equation}
m_{e_n} = \sqrt{\frac{\lambda _e ^2 (\mu ^2 - \frac{b^2}{R^2})}{2\lambda} + \frac{(n-b)^2}{R^2}}
\end{equation}
\begin{equation}
m_{u_n} = \sqrt{\frac{\lambda _u ^2 (\mu ^2 - \frac{b^2}{R^2})}{2\lambda} + \frac{(n+\frac{2}{3}b)^2}{R^2}}
\end{equation}
\begin{equation}
m_{d_n} = \sqrt{\frac{\lambda _d ^2 (\mu ^2 - \frac{b^2}{R^2})}{2\lambda} + \frac{(n+\frac{1}{3}b)^2}{R^2}}
\end{equation}
and
\begin{equation}
m_{{\nu_e}_n} = \frac{|n|}{R} \, .
\end{equation}
As in [5], the following transformations were performed on the fermion spinors to get physical mass terms for the modes:
$e_n \longrightarrow e^{i\beta _n \gamma ^5}e_n$ where 
\begin{equation}
e^{2i\beta _n \gamma ^5} = \cos{2\beta _n} + i\gamma ^5 \sin{2\beta _n} = \frac{\lambda _e v}{m_{e_n}} - i\gamma ^5 \frac{n-b}{m_{e_n}R} \nonumber
\end{equation}
as well as  $d_n \longrightarrow e^{i\sigma _n \gamma ^5}d_n$ where 
\begin{equation}
e^{2i\sigma _n \gamma ^5} = \cos{2\sigma _n} + i\gamma ^5 \sin{2\sigma _n} = \frac{\lambda _d v}{m_{d_n}} - i\gamma ^5 \frac{n+\frac{1}{3}b}{m_{d_n}R} \nonumber 
\end{equation}
and $u_n \longrightarrow e^{i\gamma _n \gamma ^5}u_n$ where 
\begin{equation}
e^{2i\gamma _n \gamma ^5} = \cos{2\gamma _n} + i\gamma ^5 \sin{2\gamma _n} = \frac{\lambda _u v}{m_{u_n}} - i\gamma ^5 \frac{n+\frac{2}{3}b}{m_{u_n}R} \nonumber
\end{equation}
and finally $ \nu_{e_n} \longrightarrow i\gamma ^5 \nu_{e_n}$ 
\section{The tunneling potential}
We form the following Lagrangian density for our Higg's field:
\begin{equation}
\EuScript{L} = (\partial _A \chi)^\dagger (\partial ^A \chi) - \mu ^2 \chi ^\dagger \chi + \frac{\lambda '}{2}(\chi ^\dagger \chi)^2  
\end{equation}
Let us then go to a Euclidean metric by letting $t \rightarrow it$. So with the flux we have \\ $\chi \underset{\rm{flux}}{\longrightarrow} \begin{pmatrix} e^{iby/R} & 0 \\ 0 & 1 \end{pmatrix} \chi$ \, or letting $B = \begin{pmatrix} e^{iby/R} & 0 \\ 0 & 1 \end{pmatrix}$,
\begin{equation} 
\chi \underset{\rm{flux}}{\longrightarrow}B \chi .
\end{equation}
Then  
\begin{equation}
\EuScript{L} = (\partial _\mu \chi)^\dagger (\partial ^\mu \chi) + \partial _y (\chi ^\dagger B^\dagger)\partial _y (B \chi) - \mu ^2 \chi ^\dagger \chi + \frac{\lambda '}{2}(\chi ^\dagger \chi)^2 \nonumber 
\end{equation}
or
\begin{equation}
\EuScript{L} = (\partial _\mu \chi)^\dagger (\partial ^\mu \chi) + \Bigg [ (\partial _y \chi ^\dagger)B^\dagger + \chi ^\dagger \begin{pmatrix} -\frac{ib}{R}e^{-iby/R} & 0 \\ 0 & 0 \end{pmatrix} \Bigg ] \Bigg [ B(\partial _y \chi) + 
\begin{pmatrix} \frac{ib}{R}e^{iby/R} & 0 \\ 0 & 0 \end{pmatrix} \chi \Bigg ] \nonumber 
\end{equation}
\begin{equation}
- \mu ^2 \chi ^\dagger \chi  + \frac{\lambda '}{2}(\chi ^\dagger \chi)^2
\end{equation}

Forming the action for the model,
\begin{equation} 
S = \int \int _0 ^{2\pi R}\EuScript{L}(x^\mu ,y) d^4 x \, dy  
\end{equation}
or 
\begin{equation}
S = \int d^4 x  \Bigg ( \sum _{n\,=\,-\infty} ^\infty \Bigg [ (\partial _\mu \chi _n)^\dagger (\partial ^\mu \chi _n) 
+ \Bigg ( -\frac{in}{R} \chi _n ^\dagger + \chi _n ^\dagger \begin{pmatrix} -\frac{ib}{R} & 0 \\ 0 & 0 \end{pmatrix} 
\Bigg ) \Bigg ( \frac{in}{R} \chi _n + \begin{pmatrix} \frac{ib}{R} & 0 \\ 0 & 0 \end{pmatrix}\chi _n \Bigg ) \nonumber 
\end{equation}
\begin{equation}
- \mu ^2 \chi _n ^\dagger \chi _n \Bigg ]+ \frac{\lambda '}{4\pi R} \sum _{n,m,l\,=\,-\infty} ^\infty 
\chi _{n-m+l} ^\dagger \chi _n \chi _m ^\dagger \chi _l \Bigg )
\end{equation}
where 
\begin{equation}
\chi (x^\mu ,y) = \begin{pmatrix} \frac{1}{\sqrt{2\pi R}}\sum _{n\,=\,-\infty} ^\infty \chi _1 (x^\mu) _n e^{iny/R} \\
\frac{1}{\sqrt{2\pi R}}\sum _{n\,=\,-\infty} ^\infty \chi _2 (x^\mu) _n e^{iny/R} \end{pmatrix}.   
\end{equation}
Since we are analyzing the tunneling for the zero modes (this procedure is not possible for arbitrary mode number because we need to use derivative techniques which are impossible with the quadric term which involves a triple sum),
\begin{equation}
S_0 = \int d^4 x \Bigg [ (\partial _\mu \chi _0)^\dagger (\partial ^\mu \chi _0) + \chi _0 ^\dagger 
\begin{pmatrix} -\mu ^2 + \frac{b^2}{R^2} & 0 \\ 0 & -\mu ^2 \end{pmatrix}\chi _0 +
\frac{\lambda '}{4\pi R}(\chi _0 ^\dagger \chi _0 )^2 \Bigg ].
\end{equation}
Let 
\begin{equation}
\chi _0 = \phi = \begin{pmatrix} \phi _1 \\ \phi _2 \end{pmatrix} = 
\begin{pmatrix} \phi _a + i\phi _b \\ \phi _c + i\phi _d \end{pmatrix}
\end{equation}
and also let $\lambda = \frac{\lambda '}{2\pi R}$. Then define 
\begin{equation}
U = -\phi ^\dagger \begin{pmatrix} \mu ^2 - \frac{b^2}{R^2} & 0 \\ 0 & \mu ^2 \end{pmatrix}\phi + \frac{\lambda}{2}(\phi ^\dagger \phi)^2
\end{equation}
or 
\begin{equation}
U = -( \mu ^2 - \frac{b^2}{R^2})|\phi _1 |^2 - \mu ^2 |\phi _2 |^2 + \frac{\lambda}{2}(|\phi _1 |^2 + |\phi _2 |^2 )^2.
\end{equation}
In minimizing the potential there is a local minimum at $\begin{pmatrix} \sqrt{\frac{\mu ^2 - \frac{b^2}{R^2}}{\lambda}} \\ 0 \end{pmatrix}$ and a global minimum at  $\begin{pmatrix} 0 \\ \sqrt{\frac{\mu ^2}{\lambda}} \end{pmatrix}$. Then the global minimum is the true vacuum and the local minimum is the false vacuum [4],
\begin{equation}
\phi _- = \begin{pmatrix} 0 \\ \sqrt{\frac{\mu ^2}{\lambda}} \end{pmatrix} 
\end{equation}
and
\begin{equation}
\phi _+ = \begin{pmatrix} \sqrt{\frac{\mu ^2 - \frac{b^2}{R^2}}{\lambda}} \\ 0 \end{pmatrix} \, .
\end{equation}
\section{Tunneling analysis}
The probability per unit time per unit volume is [4]
\begin{equation}
\frac{\Gamma}{V} = Ae^{-\frac{B}{\hbar}}\Bigg [ 1 + O(\hbar) \Bigg ] \, .
\end{equation} 
We will be calculating the parameter $B$ (sometimes called the "bounce") where here $B$ is not to be confused with the flux matrix defined before. We will not calculate the coefficient $A$ in this paper since for our numbers the tunneling probability will be completely dominated by the decaying exponential as we will see in section VI.    
The bounce is defined as [4]
\begin{equation}
B = S_0 (\phi) - S_0 (\phi _+) \,  
\end{equation}
and the difference in the energy densities of the two vacuums [4]
\begin{equation}
\epsilon = U(\phi _+ ) - U(\phi _- ) = \frac{\mu ^4}{2\lambda} - \frac{(\mu^2 - \frac{b^2}{R^2})^2 }{2\lambda}.
\end{equation}
If $t \longrightarrow it$ making the metric euclidean as we noted before, then 
\begin{equation}
(\partial _\mu \phi)^\dagger (\partial ^\mu \phi) + U(\phi ^\dagger \phi) \nonumber
\end{equation}
is 0(4) symmetric which means that $\phi (x^\mu) = \phi(\rho)$ [4] where $\rho$ is the radial vector pointing from the origin to a point in a 4-D euclidean sphere. 
From equation (34) we have \, $(\partial _\mu \phi)^\dagger (\partial ^\mu \phi) = \phi _a ^{'2} + \phi _b ^{'2} + \phi _c ^{'2} + \phi _c ^{'2}$ \, where $'$ denotes $\frac{d}{d\rho}$, then for the equations of motion
\begin{equation}
\phi '' _a  + \frac{3}{\rho}\phi ' _a = \frac{\partial U}{\partial \phi _a } 
\end{equation}
with similar equations for $\phi _b , \phi _c , \phi _d $. Then let $U(\phi) = U_0 (\phi) + O(\epsilon)$ 
($\epsilon \ll 1)$ [4] where \\
$U_0 (\phi _+ ) = U_0 (\phi _- )$ and $\frac{\partial U_0 (\phi _+ )}{\partial |\phi _1 |} = \frac{\partial U_0 (\phi _+ )}{\partial |\phi _2 |} = 0$ and $\frac{\partial U_0 (\phi _- )}{\partial |\phi _1 |} = \frac{\partial U_0 (\phi _- )}{\partial |\phi _2 |} = 0$. We find 
\begin{equation}
U_0 = -\phi ^\dagger \begin{pmatrix} \mu ^2  & 0 \\ 0 & \mu ^2 - \frac{b^2}{R^2} \end{pmatrix} \phi + \frac{\lambda}{2}
\Bigg [ \phi ^\dagger \begin{pmatrix} \sqrt{\frac{\mu ^2}{\mu ^2 - \frac{b^2}{R^2}}}  & 0 \\ 0 & \sqrt{\frac{\mu ^2 - \frac{b^2}{R^2}}{\mu ^2}} \end{pmatrix} \phi \Bigg ] ^2 
\end{equation}
or
\begin{equation}
U_0 = -\mu ^2 |\phi _1 |^2 - (\mu ^2 - \frac{b^2}{R^2})|\phi _2 |^2 + \frac{\lambda}{2}\Bigg ( \sqrt{\frac{\mu ^2}{\mu ^2 - \frac{b^2}{R^2}}} \, |\phi _1 |^2 + \sqrt{\frac{\mu ^2 - \frac{b^2}{R^2}}{\mu ^2}} \, |\phi _2 |^2 \Bigg ) ^2.
\end{equation}

The second term in equation (42) can be neglected [4], this will be justified later. Then
\begin{equation}
\phi '' _a \approx \frac{\partial U_0 }{\partial \phi _a } 
\end{equation}
and similar equations for $\phi _b , \phi _c , \phi _d $. Equation (45) can be written as 
\begin{equation}
\phi ' _a \, d\phi ' _a = \frac{\partial U_0 }{\partial \phi _a }\, d\phi _a \nonumber 
\end{equation}
and similarly for $\phi _b , \phi _c , \phi _d $. Then we can write
\begin{equation}
\phi ' _a \, d\phi ' _a + \phi ' _b \, d\phi ' _b + \phi ' _c \, d\phi ' _c + \phi ' _d \, d\phi ' _d = 
\frac{\partial U_0 }{\partial \phi _a }\, d\phi _a + \frac{\partial U_0 }{\partial \phi _b }\, d\phi _b + 
\frac{\partial U_0 }{\partial \phi _c }\, d\phi _c + \frac{\partial U_0 }{\partial \phi _d }\, d\phi _d \nonumber 
\end{equation}
or since 
\begin{equation}
\frac{\partial U_0 }{\partial \phi _a }\, d\phi _a + \frac{\partial U_0 }{\partial \phi _b }\, d\phi _b + 
\frac{\partial U_0 }{\partial \phi _c }\, d\phi _c + \frac{\partial U_0 }{\partial \phi _d }\, d\phi _d = d U_0 \nonumber
\end{equation}
which gives finally,  
\begin{equation}
\frac{1}{2} \phi _a ^{'2} + \frac{1}{2} \phi _b ^{'2} + \frac{1}{2} \phi _c ^{'2} + \frac{1}{2} \phi _d ^{'2} - U_0 = 
-U_0 (\phi _+ ).
\end{equation}
The constant of integration was determined by the boundary condition $\phi (\rho \rightarrow \infty ) = \phi _+ $. 
This can be seen by imagining $B$ as the true vacuum states enclosed by a 4-D sphere of radius $\bar{\rho}$ [4] where outside the sphere wall there is the false vacuum states [4]. This picture is accurate for large $\bar{\rho}$ where $\bar{\rho}$ is the point where $\phi$ is the average of the two extreme values $\phi _+ $ and $\phi _- $ [4]. With this picture then
\begin{equation} 
B = B_{\rm{outside}} + B_{\rm{wall}} + B_{\rm{inside}}. 
\end{equation}
So now we can easily see that $B_{\rm{outside}} = 0$ from equation (40). Also 
\begin{equation}
B_{\rm{inside}} = 2\pi ^2 \int _0 ^\infty d\rho \, \rho ^3 \, \Bigg [ U_0 (\phi _- ) - U_0 (\phi _+ )\Bigg ]\approx
-\frac{\pi ^2}{2}\bar{\rho}^4 \epsilon
\end{equation}
where remember that $\epsilon = U(\phi _+ ) - U(\phi _- )$ and we are assuming that $\bar{\rho}$ is large so that 
$\int _0 ^\infty d\rho \approx \int _0 ^{\bar{\rho}}d\rho$. In the thin wall approximation
\begin{equation}
B_{\rm{wall}} = 2\pi ^2 \bar{\rho}^3 \int _0 ^\infty d\rho \, \Bigg [ \frac{1}{2} \phi _a ^{'2} + \frac{1}{2} \phi _b ^{'2} + \frac{1}{2} \phi _c ^{'2} + \frac{1}{2} \phi _d ^{'2} + U_0 (\phi) - U_0(\phi _+)\Bigg ]
\end{equation} 
or using equation (46) we can write [4]
\begin{equation}
B_{\rm{wall}} = 2\pi ^2 \bar{\rho}^3 \int _0 ^\infty d\rho \, 2 \Bigg [ U_0 - U_0 (\phi _+ )\Bigg ] 
\end{equation}
and let
\begin{equation}
S_1 = \int _0 ^\infty d\rho \, 2 \Bigg [ U_0 - U_0 (\phi _+ )\Bigg ].
\end{equation}
Then we have 
\begin{equation}
B = B_{\rm{outside}} + B_{\rm{wall}} + B_{\rm{inside}} = -\frac{\pi ^2}{2}\bar{\rho}^4 \epsilon + 2\pi ^2 \bar{\rho}^3 S_1.
\end{equation}
$B$ should be stationary under variations of $\bar{\rho}$ [4], so $B$ is stationary at 
$\bar{\rho} = \frac{3 S_1}{\epsilon}$. It should be pointed out that $\bar{\rho}$ is on the order of 10GeV in using the upper limit of $\frac{|b|}{R}$ = 3GeV [5] in the expressions for $S_1$ (see equation (60)) and $\epsilon$. Thus it is necessary to use a much smaller $\frac{|b|}{R}$ than the upper limit to justify neglecting the second term in (42) as stated earlier [4]. Then 
\begin{equation}
B = \frac{27\pi ^2 S_1 ^4}{2\epsilon ^3}.
\end{equation}
and all that remains is to determine $S_1$. Recall that the probability per unit time per unit volume is [4]
\begin{equation}
\frac{\Gamma}{V} = Ae^{-\frac{B}{\hbar}}\Bigg [ 1 + O(\hbar) \Bigg ] \, . \nonumber
\end{equation} 

Let us now tackle $S_1$. First, we go back to the equation of motion for $\phi _a$ which gives 
\begin{equation}
\frac{1}{2}\phi _a ^{'2} = \int \frac{\partial U_0}{\partial \phi _a}d\phi _a \,+ \,\rm{constant} 
\end{equation}
and since $\phi (\rho \rightarrow \infty ) = \phi _+ $ we must have 
\begin{equation}
\frac{1}{2}\phi _a ^{'2} = \int \frac{\partial U_0 (\phi _a, \phi _b = 0,  \phi _c = 0, \phi _d = 0)}{\partial \phi _a}d\phi _a \, + \, \frac{\mu ^2 (\mu ^2 - \frac{b^2}{R^2})}{2\lambda}. 
\end{equation}
To see this a little more clearly, we note that equation (54) must hold in general for all $\rho$. $\phi _b , \phi _c ,$ and $\phi _d$ are all constants with respect to the integrals on both sides of (54) (they are not constants with respect to $\rho$ obviously). Then with the boundary condition $\phi (\rho \rightarrow \infty ) = \phi _+ $, the LHS vanishes and so must the RHS when evaluated at $\phi _+ $. Therefore, as constants of integration in (54), $\phi _b , \phi _c ,$ and $\phi _d$ are completely determined  by the boundary condition. Thus
\begin{equation}
\phi ' _a = \sqrt{2}\sqrt{ -\mu ^2 \phi _a ^2 + \frac{\lambda}{2}\frac{\mu ^2}{\mu ^2 - \frac{b^2}{R^2}}\phi _a ^4
+ \frac{\mu ^2 (\mu ^2 - \frac{b^2}{R^2})}{2\lambda}}\, .
\end{equation}
Similarly,
\begin{equation}
\phi ' _b = \sqrt{\frac{\lambda \mu ^2}{\mu ^2 - \frac{b^2}{R^2}}}\, \phi _b ^2 \, , 
\end{equation}
\begin{equation}
\phi ' _c = \sqrt{\frac{\lambda (\mu ^2 - \frac{b^2}{R^2})}{\mu ^2}}\, \phi _c ^2 \, , 
\end{equation}
and
\begin{equation}
\phi ' _d = \sqrt{\frac{\lambda (\mu ^2 - \frac{b^2}{R^2})}{\mu ^2}}\, \phi _d ^2 \, . 
\end{equation}

Then 
\begin{equation}
d\rho = \frac{\partial \rho}{\partial \phi _a}d\phi _a + \frac{\partial \rho}{\partial \phi _b}d\phi _b + 
\frac{\partial \rho}{\partial \phi _c}d\phi _c + \frac{\partial \rho}{\partial \phi _d}d\phi _d \nonumber \, = \, \frac{d\phi _a}{\phi ' _a} + \frac{d\phi _b}{\phi ' _b} + \frac{d\phi _c}{\phi ' _c} + \frac{d\phi _d}{\phi ' _d} \nonumber
\end{equation}
where we will have to be careful to evaluate the constants for each integrand at their corresponding $\phi _+$ values (in the numerator of the integrand for $S_1$). Then
\begin{equation}
S_1 = 2\int _{\phi _{- _a}} ^{\phi _{+ _a}}\frac{U_0 (\phi _a ,0,0,0) - U_0 (\phi _+ )}{\phi ' _a (\phi _a ,0,0,0)}d\phi _a +  2\int _{\phi _{- _b}} ^{\phi _{+ _b}}\frac{U_0 (\phi _a=\phi _{+ _a} ,\phi _b ,0,0) - U_0 (\phi _+ )}{\phi ' _b (\phi _a=\phi _{+ _a} ,\phi _b ,0,0)}d\phi _b \nonumber
\end{equation}
\begin{equation}
+ 2\int _{\phi _{- _c}} ^{\phi _{+ _c}}\frac{U_0 (\phi _a=\phi _{+ _a} ,0,\phi _c ,0) - U_0 (\phi _+ )}{\phi ' _c (\phi _a=\phi _{+ _a} ,0,\phi _c ,0)}d\phi _c +  2\int _{\phi _{- _d}} ^{\phi _{+ _d}}\frac{U_0 (\phi _a=\phi _{+ _a} ,0 ,0,\phi _d) - U_0 (\phi _+ )}{\phi ' _d (\phi _a=\phi _{+ _a} ,0 ,0,\phi _d)}d\phi _d \nonumber
\end{equation}
or
\begin{equation}
S_1 = \sqrt{2}\int _0 ^{\sqrt{\frac{\mu ^2 - \frac{b^2}{R^2}}{\lambda}}}\sqrt{\frac{\lambda}{2}\frac{\mu ^2}{\mu ^2 - \frac{b^2}{R^2}}\phi _a ^4 - \mu ^2 \phi _a ^2 + \frac{\mu ^2 (\mu ^2 - \frac{b^2}{R^2})}{2\lambda}} \, d\phi _a \,
+ \, \sqrt{\frac{\lambda (\mu ^2 - \frac{b^2}{R^2})}{\mu ^2}}\int _{\sqrt{\frac{\mu ^2}{\lambda}}} ^0 \phi _c ^2 \, d\phi _c \nonumber
\end{equation}
where the $\phi _b$ and $\phi _d$ integrals vanish. The $\phi _a$ integral involves several elliptic integrals but nonetheless this expression vanishes when evaluated at the integral limits. Thus the only non vanishing integral is for $\phi _c$ which gives
\begin{equation}
\sqrt{\frac{\lambda (\mu ^2 - \frac{b^2}{R^2})}{\mu ^2}}\int _{\sqrt{\frac{\mu ^2}{\lambda}}} ^0 \phi _c ^2 \, d\phi _c 
\, = \, - \, \frac{\mu ^2 \sqrt{\mu ^2 - \frac{b^2}{R^2}}}{3\lambda}\, . \nonumber
\end{equation}
Therefore 
\begin{equation}
S_1 = - \, \frac{\mu ^2 \sqrt{\mu ^2 - \frac{b^2}{R^2}}}{3\lambda}\, .
\end{equation} 
\section{Tunneling probability result}
We have from equation (53) finally,
\begin{equation}
B = \frac{4\pi ^2}{3\lambda} \, \frac{\mu ^8 (\mu ^2 - \frac{b^2}{R^2})^2}{(\mu ^4 - (\mu ^2 - \frac{b^2}{R^2})^2)^3}\, .
\end{equation}
If we do a very simple calculation of $B$ using a Higgs mass = 2$\mu$ of 150GeV along with the vacuum $v = \sqrt{\frac{\mu^2}{\lambda}}$ of 250GeV. Then we find $\mu$ = 75GeV and $\lambda$ = 0.09 (unitless). Finally with  $\frac{|b|}{R}$ = 3GeV we find $B = 4.5\rm{x}10^9$. The value of $B$ gets larger for smaller $\frac{|b|}{R}$ than 3GeV. Thus the lifetime for tunneling from the false vacuum to the true vacuum is vanishingly small and completely dominated by the decaying exponential. In the limit as $b \rightarrow 0$ then $\phi _+ = \phi _-$ and the vacuum becomes degenerate. In this limit we should have $\frac{\Gamma}{V} \rightarrow 0$ for the tunneling probability (from our interpretation of the tunneling probability). From equation (61), $B \rightarrow \infty$ in the limit as $b$ goes to zero and thus from (39) we see that indeed 
\begin{equation}
\frac{\Gamma}{V} \underset{b\,=\,0}{\rightarrow} 0 \nonumber
\end{equation}
as expected. We are not concerned with the coefficient $A$ in (39) as this constant, to my knowledge, can only be determined numerically [4]. This constant obviously does not affect the overall trend of the tunneling probability who's behavior is completely dominated by the decaying exponential for our numbers.

\end{document}